\documentclass[pre,twocolumn,showpacs]{revtex4}
\usepackage{amsmath,amssymb,mathrsfs}
\usepackage{graphicx}
\usepackage{graphics}
\usepackage{epsfig}
\usepackage{bm}
\usepackage{color}
\usepackage{verbatim,color,ulem}

\begin{document}

\title{Emulation of lossless exciton-polariton condensates by dual-core
optical waveguides: Stability, collective modes, and dark solitons}
\author{Luca Salasnich$^{1,2}$, Boris A. Malomed$^{3}$, and Flavio Toigo$%
^{1} $}
\affiliation{$^1$Dipartimento di Fisica e Astronomia 
``Galileo Galilei'' and CNISM,
Universit\`a di Padova, Via Marzolo 8, 35131 Padova, Italy \\
$^{2}$Istituto Nazionale di Ottica (INO) del Consiglio Nazionale delle
Ricerche (CNR), Sezione di Sesto Fiorentino, Via Nello Carrara, 1 - 50019
Sesto Fiorentino, Italy \\
$^{3}$Department of Interdisciplinary Studies, School of Electrical
Engineering, Faculty of Engineering, Tel Aviv University, Tel Aviv 69978,
Israel}

\begin{abstract}
We propose a possibility to simulate the exciton-polariton (EP)
system in the lossless limit, which is not currently available in
semiconductor microcavities, by means of a simple optical dual-core
waveguide, with one core carrying the nonlinearity and operating close to
the zero-group-velocity-dispersion (GVD)\ point, and the other core being
linear and dispersive. Both 2D and 1D EP systems may be emulated by means of
this optical setting. In the framework of this system, we find that, while
the uniform state corresponding to the lower branch of the nonlinear
dispersion relation is stable against small perturbations, the upper branch
is always subject to the modulational instability (MI). The stability and
instability are verified by direct simulations too. We analyze collective
excitations on top of the stable lower-branch state, which include a
Bogoliubov-like gapless mode and a gapped one. Analytical results are
obtained for the corresponding sound velocity and energy gap. The effect of
a uniform phase gradient (superflow) on the stability is considered too,
with a conclusion that the lower-branch state becomes unstable above a
critical wavenumber of the flux. Finally, we demonstrate that the stable 1D
state may carry robust dark solitons.
\end{abstract}

\maketitle

\section{Introduction and the objective}

\textit{Emulation} of complex effects and systems known in condensed-matter
physics by means of simpler and ``cleaner" settings, based on classical
photonic, or quantum-mechanical atomic, waves has recently drawn a lot of
interest \cite{emulator}. The first example is provided by the
superfluidity, which may be studied in a much more accurate form in atomic
Bose-Einstein condensates (BECs) \cite{Pitaevskii} and ultracold Fermi gases
\cite{Fermi} than in liquid helium. Another possibility, which has come to
the forefront recently, is the experimental realization \cite{Nature} and
theoretical analysis \cite{Rashba-BEC} of the (pseudo-) spin-orbit coupling
in a binary BEC, induced by specially designed laser fields, see a brief
review of the topic in Ref. \cite{Zhai}. Similar techniques were recently
developed for the creation of synthetic Abelian and non-Abelian gauges
fields in atomic BEC \cite{Abelian,review}. As concerns photonics, it is
well known that it allows an efficient experimental emulation of
fundamentally important settings known in condensed matter, such as the
Anderson localization \cite{Moti-And} (the experimental realization of this
effect in BEC has been demonstrated too \cite{BEC-And}), graphene \cite%
{Moti-graph}, and topological insulators \cite{Moti-top-ins}. Furthermore,
the use of the wave propagation in photonic media opens the way for
experimental simulation, in terms of classical physics, of fundamental
phenomena predicted in the quantum theory, which are very difficult to
observe directly, such as non-Hermitian Hamiltonians which generate real
spectra due to the $\mathcal{PT}$ symmetry \cite{PT,PT2}, and exotic
relativistic effects (\textit{Zitterbewegung} and others) \cite%
{Longhi,Longhi2}.

The photonic and matter-wave systems may often be used to emulate each
other. For instance, the system of coupled Gross-Pitaevskii equations (GPEs)
realizing the spin-orbit coupling in the 1D setting \cite{Kevrekidis} is
exactly tantamount to the earlier studied system of coupled nonlinear Schr%
\"{o}dinger equations (NLSEs) modeling a twisted bimodal optical fiber \cite%
{me}, making solitons in these systems also mutually equivalent.

The emulation methods offer an additional advantage, making it possible to
attain physical conditions and effects in simulating systems which are
inaccessible in the original ones. An obvious example is provided by
matter-wave solitons, which can be readily created in rarefied atomic gases,
cooled into the BEC state \cite{solitons}, while they are not observed in
dense superfluids.

Another important topic, combining semiconductor physics and photonics,
which has recently drawn \ a great deal of interest, is the strong coupling
of light (cavity photons) and matter (excitons, i.e., bound electron-hole
states) in semiconductor microcavities \cite{deveaud}. 
It is well established that this
interaction leads to the creation of hybrid modes in the form of
exciton-polaritons (EPs) \cite{KBM+2007,rev-francesca,rev-iacopo}. The EP
nonlinearity is self-defocusing due to the electrostatic repulsion between
excitons. The nonlinearity plays an important role in a number of effects
predicted and (partly) observed in EP systems, such as bistability \cite%
{BKE+2004,gip,CC2004}, wave mixing \cite{gip,CC2004,SBS+2000}, superfluidity
\cite{CC2004,bogol} and the formation of dark and bright solitons \cite%
{agr,skryabin1,skryabin2,Kivshar,soliton-first,recent}, as well as of gap
solitons (of the bright type), produced by the interplay of the
self-repulsive nonlinearity with a spatially periodic linear potential \cite%
{skryabin3}. The nonlinearity manifests itself too in EP bosonic
condensates, which have been created in the experiment \cite{exp1}, and used
to demonstrate Bogoliubov excitations \cite{exp2}, diffusionless motion \cite%
{exp3}, persistent currents and quantized vortices \cite{exp4}, among other
effects.

In experiments based on incoherent pumping \cite{exp1,exp2,exp3},
off-resonance pumped polaritons scatter down, loosing the coherence
inherited from the pump, and go into the condensate which emerges at the
lower branch of the EP dispersion law. Real EP condensates are very well
described by the extended GPE which takes into account the pump and loss
\cite{rev-iacopo,exp1,berloff,carusotto}. A more general approach adopts a
system of two Rabi- (linearly) coupled equations, \textit{viz}., the GPE for
the wave function (order parameter) of excitons, and the propagation
equation of the linear-Schr\"{o}dinger type for the amplitude of the
cavity-photon field \cite{ciuti}. In particular, these equations have been
used to predict the existence of the above-mentioned dark \cite{skryabin1},
bright \cite{skryabin2} and gap \cite{skryabin3} EP solitons. The same
equations have been used to investigate the stability of the EP fluid under
coherent pumping \cite{fran1}.

In most cases, the Rabi-coupled system includes the loss and pump
terms in the exciton and photon-propagation equations, respectively.
In some works, it was assumed that the system maintains the
background balance between the pump and loss in the first
approximation, allowing one to consider effectively lossless
dynamics \cite{agr,soliton-first,recent,skryabin3}. While this
``ideal" version of the EP model makes it possible to predict a
number of potentially interesting effects, such as solitons, in a
relatively simple form, it is not realistic for the description of
the EP dynamics in semiconductor cavities. This problem
suggests to look for feasible photonic systems which would be able to \emph{%
emulate} the lossless version of the EP model. In fact, such photonic
systems were proposed, without and relation to EP models, in Refs. \cite%
{Javid} and \cite{Arik}. They are based on asymmetric dual-core optical
fibers (or a photonic-crystal fibers with two embedded cores \cite%
{dual-core-PCF}), with the linear coupling between the cores emulating the
Rabi coupling between excitons and cavity photons in the EP system. It is
assumed that only one core is nonlinear (which can be easily realized by
engineering an appropriate transverse modal structure or using
nonlinearity-enhancing dopants \cite{Lenstra}), operating close to the
zero-dispersion point, while in the mate (linear) core the group-velocity
dispersion (GVD)\ is normal or anomalous \cite{Arik}, if the nonlinearity
sign in the first core is self-focusing or defocusing, respectively.
Alternatively, the linear core may carry a Bragg grating \cite{Javid}, which
offers the optical emulation of the model for the EP gap solitons introduced
in Ref. \cite{skryabin3}.

It is relevant to mention that the EP system in semiconductor microcavities
may be excited solely by the pump injecting cavity photons. The emulation
scheme based on the similarity to the dual-core optical fiber opens an
additional possibility, to excite various states in the system by injecting
the field into the nonlinear core, which simulates the excitonic wave
function.

The above-mentioned temporal-domain dual-fiber-based setting, which was
introduced in Ref. \cite{Arik}, is exactly tantamount to the lossless limit
of the one-dimensional (1D) EP system, see Eqs. (\ref{eq1pip}) and (\ref%
{eq2pip}) below (in the optical fiber, the losses may be easily kept
negligible for an experimentally relevant propagation distance, or, if
necessary, compensated by built-in gain). The optical emulation of the 2D
version of the EP system is more tricky, but possible too. In the latter
case, one may introduce the system of \ spatiotemporal NLSEs for the
dual-core planar waveguide. The 2D diffraction of cavity photons is then
emulated by the combination of the transverse diffraction and anomalous GVD
in the linear core. Accordingly, the nonlinearity in the mate core, kept
near the zero-GVD point, must be self-defocusing. Furthermore, to suppress
the transverse diffraction in the nonlinear core (to emulate the
non-existing or very weak diffraction of excitons), the nonlinear core
should be built as an array of fibers, rather than as a solid waveguide.
This 2D setting which emulates the lossless EP system is based on Eqs. (\ref%
{eq1}) and (\ref{eq2}) presented below. Such a planar dual-core waveguide,
in which one core is solid, while the other one is represented by an array
of 1D waveguides, is quite possible \cite{Nicolae}.

Using the emulating counterpart of the lossless EP system, we address new
possibilities suggested by this emulation . In particular, we consider the
dynamics on the upper branch of the nonlinear dispersion relation, which are
usually disregarded in the dissipative EP system. The issues addressed below
include the modulational instability (MI)\ of uniform states corresponding
to the upper and lower branches and collective excitations on top of the
stable background (various forms of the dispersion relation and excitations
on top of the lower polariton branch in dissipative EP systems were studied
earlier \cite{dispersion}). In the case when the uniform background is
stable, we consider dark solitons too (in the dissipative EP model, such
solitons were recently studied in Ref. \cite{dark-EP}.

The rest of the paper is structured as follows. The 2D system, which is
emulated, as said above, by the planar dual-core waveguide in the
spatiotemporal domain, is introduced in Section II. The MI and collective
excitations on top of the stable (lower) branch of the nonlinear dispersion
relation are considered in Section III. Effects of the phase gradient on the
stability are investigated in Section IV. The reduction of the 2D model to
1D, and the investigation of dark solitons in the latter case, are presented
in Section V. The paper is concluded by Section VI.

\section{The model}

The spatiotemporal evolution of complex amplitudes of the electromagnetic
field in the nonlinear and linear cores of the planar waveguide, $\psi $ and
$\phi $, obeys the system of coupled NLSEs \cite{Arik}, which is written
here in the notation corresponding to the emulation of the EP system 
\cite{deveaud,KBM+2007,rev-francesca,rev-iacopo} by means of the optical model:
\begin{eqnarray}
i{\frac{\partial }{\partial t}}{\psi } &=&\left[ -{\frac{1}{2m_{X}}}\left(
\frac{\partial ^{2}}{\partial x^{2}}+\frac{\partial ^{2}}{\partial y^{2}}%
\right) +\epsilon _{X}+g\,|{\psi }|^{2}\right] {\psi }+\Gamma \ {\phi }\;,
\label{eq1} \\
i{\frac{\partial }{\partial t}}{\phi } &=&\left[ -{\frac{1}{2m_{C}}}\left(
\frac{\partial ^{2}}{\partial x^{2}}+\frac{\partial ^{2}}{\partial y^{2}}%
\right) +\epsilon _{C}\right] {\phi }+\Gamma \ {\psi }\;,  \label{eq2}
\end{eqnarray}%
where $t$ (corresponding to time in the EP system) is the propagation
distance along the waveguide, $x$ and $y$ are, respectively, the transverse
coordinate and reduced time, both corresponding to spatial coordinates in
the emulated semiconductor microcavity, while $m_{X}$ and $m_{C}$, which 
correspond to the effective excitonic and cavity-photon masses, are actually
the inverse diffraction-dispersion coefficients in the two cores 
\cite{deveaud,KBM+2007,rev-francesca,rev-iacopo}. The EP
setting typically has $m_{C}/m_{X}\sim 10^{-4}$ 
\cite{rev-francesca,rev-iacopo}, which implies that, as said above, 
the nonlinear 
core of the waveguide operates very close to the zero-GVD point, while the
diffraction is suppressed by the fact that this core is built as an array
of 1D waveguides [it is assumed that the small residual GVD\ and diffraction
in the nonlinear core are adjusted so as not to break the spatiotemporal
isotropy of the optical system\ in the $\left( x,y\right) $ plane]. Further,
$\epsilon _{X}$ and $\epsilon _{C}$ are propagation-constant shifts in the
two waveguides, which, in terms of the EP, represent, respectively, the
chemical potential of excitons and photon energy at zero wavenumber.
Coefficient $g>0$ in Eq. (\ref{eq1}) represents the self-defocusing optical
nonlinearity, which corresponds to the strength of the repulsive excitonic
self-interaction. Lastly, the inter-core coupling constant, $\Gamma $,
emulates the strength of the EP Rabi coupling.

The total energy of the optical signal, which represents the number of
condensed polaritons, i.e., the sum of numbers ${N}_{X}$ and $N_{C}$ of the
excitons and photons, is
\begin{equation}
N_{0}=N_{X}+N_{C}=\int d^{2}\mathbf{r}\ \left[ |\psi (\mathbf{r}%
,t)|^{2}+|\phi (\mathbf{r},t)|^{2}\right] ,  \label{N0}
\end{equation}%
is the dynamical invariant of the lossless system. Equations (\ref{eq1}) and
(\ref{eq2}) also conserve the Hamiltonian,
\begin{eqnarray}
H &=&\int d^{2}\mathbf{r}\Big[\frac{1}{2m_{X}}\left\vert \nabla _{\perp
}\psi \right\vert ^{2}+\frac{1}{2m_{C}}\left\vert \nabla _{\perp }\phi
\right\vert ^{2} \\
&+&\epsilon _{X}\,|{\psi }|^{2}+\epsilon _{C}|\phi |^{2}+\frac{g}{2}|\psi
|^{4}+\Gamma \left( \psi ^{\ast }\phi +\psi \phi ^{\ast }\right) \Big]\;,
\notag
\end{eqnarray}%
as well as the total 2D momentum and angular momentum.

\section{Uniform states}

\subsection{Stationary uniform states}

Equations ( (\ref{eq1}) and (\ref{eq2}) give rise to an optical continuous
wave in the dual-core waveguide with propagation constant $-\mu $,
\begin{equation}
\psi (\mathbf{r},t)=\psi _{0}\ e^{-i\mu t},~\phi (\mathbf{r},t)=\phi _{0}\
e^{-i\mu t},  \label{mu}
\end{equation}
\begin{eqnarray}
\psi _{0}^{2} &=&{\frac{1}{g}}\left( \mu -\epsilon _{X}+{\frac{\Gamma ^{2}}{%
\epsilon _{C}-\mu }}\right) \;,  \label{cond1} \\
\phi _{0} &=&-{\frac{\Gamma }{(\epsilon _{C}-\mu )}}\,\psi _{0}\;,
\label{cond2}
\end{eqnarray}%
which emulates the EP\ condensate with chemical potential $\mu $ and exciton
and cavity-photon amplitudes $\psi _{0}$ and $\phi _{0}$. Then, one can
eliminate $\mu $ from Eqs. (\ref{cond1}) and (\ref{cond2}) in favor of $\psi
_{0}$:
\begin{equation}
\mu ^{\pm }={\frac{1}{2}}\Big(\epsilon _{X}+\epsilon _{C}+g\psi _{0}^{2}\pm
\sqrt{(\epsilon _{X}-\epsilon _{C}+g\psi _{0}^{2})^{2}+4\Gamma ^{2}}\Big)\;.
\label{echem}
\end{equation}%
In terms of EP, this relation includes the \textit{lower} $\mu ^{-}$ and the
\textit{upper }branches, $\mu ^{-}$ and $\mu ^{+}$, respectively.

As said above, the repulsive excitonic nonlinearity corresponds to $g>0$,
and the physically relevant EP\ setting has $\epsilon _{X}<\epsilon _{C}$
\cite{rev-francesca,rev-iacopo}. It follows from here that the uniform
EP-emulating configuration exists [i.e., Eqs. (\ref{cond1}) and (\ref{cond2}%
) yield $\psi _{0}^{2}$, $\phi _{0}^{2}>0$] provided that the chemical
potential satisfies conditions $\mu _{0}^{-}<\mu <\epsilon _{C}$ or $\mu
>\mu _{0}^{+}$, at lower the upper branch, respectively. Here, $\mu _{0}^{-}$
and $\mu _{0}^{+}$ are obtained from Eq. (\ref{echem}) by setting $\psi
_{0}=0$:
\begin{equation}
\mu _{0}^{\pm }=\frac{1}{2}\left( \epsilon _{X}+\epsilon _{C}\pm \sqrt{%
(\epsilon _{C}-\epsilon _{X})^{2}+4\Gamma ^{2}}\right) .  \label{mu0}
\end{equation}%
Thus, for $\Gamma =0$ one has $\mu _{0}^{-}=\epsilon _{X}$ and $\mu
_{0}^{+}=\epsilon _{C}$, while for $\Gamma \neq 0$ the relevant ranges are $%
\mu _{0}^{-}<\epsilon _{X}$ and $\mu _{0}^{+}>\epsilon _{C}$. The respective
EP condensate density $n_{0}$ (alias the energy density of the optical
signal in the dual-core waveguide) is
\begin{equation}
n_{0}=\psi _{0}^{2}+\phi _{0}^{2}~,  \label{tot-cond}
\end{equation}%
cf. Eq. (\ref{N0}). Note that, due to Eq. (\ref{cond2}), $\psi _{0}$ and $%
\phi _{0}$ have opposite signs on the lower branch, while on the upper one
the signs of $\psi _{0}$ and $\phi _{0}$ are identical.

\begin{figure}[t]
\centerline{\epsfig{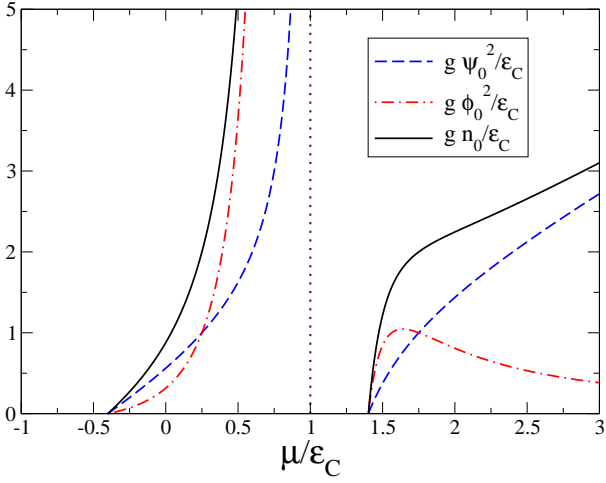}}
\caption{(Color online). Scaled effective densities of the uniform
exciton-polariton condensate versus the scaled chemical potential, $\protect%
\mu /\protect\epsilon _{C}$, for the effective (emulated) Rabi coupling $%
\Gamma =0.75\protect\epsilon _{C}$. Dashed lines: effective exciton 
density $\protect\psi _{0}^{2}$; dotted-dashed lines: the respective
cavity-photon density $\protect\phi _{0}^{2}$; solid lines: total density $%
n_{0}$. Other parameters are the effective exciton-exciton repulsion
strength $g$, and the exciton and cavity-photon energies, $\protect\epsilon %
_{X}=0$ and $\protect\epsilon _{C}$ (actually emulated by the
propagating-constant shifts in the dual-core waveguide), at zero wavenumber.
The curves below and above $\protect\mu /\protect\epsilon _{C}=1$
correspond, respectively, to the lower and upper branches.}
\label{fig1}
\end{figure}

In the framework of the present model, for given $\phi _{0}^{2}$ one can
easily obtain the effective exciton and total densities, $\psi _{0}^{2}$ and
$\phi _{0}^{2}$, along with the chemical potential, $\mu $, from from Eqs. (%
\ref{cond1}), (\ref{cond2}), and (\ref{echem}). In Fig. \ref{fig1} we
display $\psi _{0}^{2}$ (dashed lines), $\phi _{0}^{2}$ (dotted-dashed
lines), and $n_{0}$ (solid lines) as functions of the scaled chemical
potential, $\mu /\epsilon _{C}$, for $\epsilon _{X}=0$ and a relevant value
of the linear-coupling strength, $\Gamma =0.75\epsilon _{C}$. As previously
stated, the curves in the range of $\mu _{0}^{-}/\epsilon _{C}=-0.38<\mu
/\epsilon _{C}<1$ correspond to the lower branch, while the range of $\mu
/\epsilon _{C}>\mu _{0}^{+}/\epsilon _{C}=1.4$ pertains to the upper one.

\begin{figure}[t]
\centerline{\epsfig{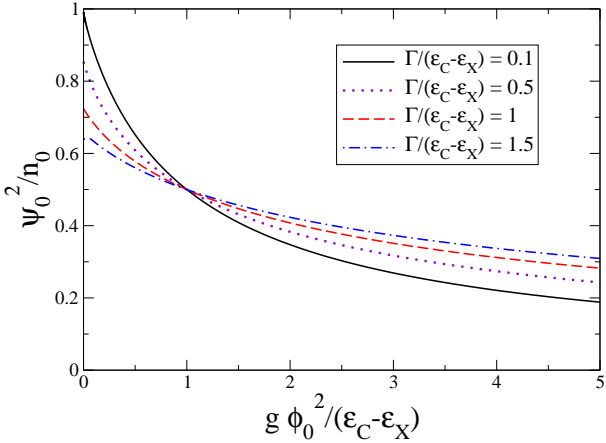}}
\caption{(Color online). Exciton fraction $\protect\psi _{0}^{2}/n_{0}$ in
the exciton-polariton condensate as a function of the scaled density of the
cavity photons, $g\protect\phi _{0}^{2}/(\protect\epsilon _{C}-\protect%
\epsilon _{X})$. The four curves correspond to different values of the
scaled Rabi coupling, $\Gamma /(\protect\epsilon _{C}-\protect\epsilon _{X})$%
. }
\label{fig2}
\end{figure}

\subsection{Modulational instability (MI)\ of the uniform states}

A central point of the analysis is the MI of the flat state which was
obtained above. For this purpose, small perturbations ${\eta }_{X}(\mathbf{r}%
,t)$ and ${\eta }_{C}(\mathbf{r},t)$ are added to the uniform fields, $\psi
_{0}$ and $\phi _{0}$, by setting
\begin{equation}
\left\{ {\psi }(\mathbf{r},t),{\phi }(\mathbf{r},t)\right\} =\left[ \left\{
\psi _{0},\phi _{0}\right\} +\left\{ {\eta }_{1}(\mathbf{r},t),{\eta }_{2}(%
\mathbf{r},t)\right\} \right] \ e^{-i\mu t}\;.
\end{equation}%
The subsequent linearization of Eqs. (\ref{eq1}) and (\ref{eq2}) gives
\begin{eqnarray}
i{\frac{\partial }{\partial t}}{\eta }_{1} &=&\left( -{\frac{1}{2m_{X}}}%
\nabla _{\bot }^{2}+\mu _{X}-2{\frac{\Gamma ^{2}}{\mu _{C}}}\right) {\eta }%
_{1}  \notag \\
&+&(\mu _{X}-{\frac{\Gamma ^{2}}{\mu _{C}}})\ {\eta }_{1}^{\ast }+\Gamma {%
\eta }_{2}\;,  \notag \\
i{\frac{\partial }{\partial t}}{\eta }_{2} &=&\left( -{\frac{1}{2m_{C}}}%
\nabla _{\bot }^{2}-\mu _{C}\right) {\eta }_{2}+\Gamma {\eta }_{1}\;,
\label{linear}
\end{eqnarray}%
where we have defined
\begin{equation}
\mu _{X}\equiv \mu -\epsilon _{X},\quad \quad \mu _{C}\equiv \mu -\epsilon
_{C}.  \label{mumumumu}
\end{equation}%
Solution to linearized equations (\ref{linear}) are looked for as
\begin{eqnarray}
\eta _{1}(\mathbf{r},t) &=&A_{1}\ e^{i(\mathbf{k}\cdot \mathbf{r}-\omega
_{k}t)}+B_{1}\ e^{-i(\mathbf{k}\cdot \mathbf{r}-\omega _{k}t)}\;,
\label{pert1} \\
\eta _{2}(\mathbf{r},t) &=&A_{2}\ e^{i(\mathbf{k}\cdot \mathbf{r}-\omega
_{k}t)}+B_{2}\ e^{-i(\mathbf{k}\cdot \mathbf{r}-\omega _{k}t)}\;,
\label{pert2}
\end{eqnarray}%
where $\mathbf{k}$ and $\omega _{k}$ are the wave vector and frequency of
the perturbations. It is straightforward to derive a dispersion relation
from Eqs. (\ref{pert1}) and (\ref{pert2}):
\begin{equation}
\omega _{k}^{\pm }=\sqrt{-\beta \pm \sqrt{\beta ^{2}-4\gamma }},
\label{freq}
\end{equation}%
where additional combinations are defined:%
\begin{eqnarray}
\beta &\equiv &-(a^{2}-b^{2}+c^{2}+2\Gamma ^{2}),~ \\
\gamma &\equiv &a^{2}c^{2}-b^{2}c^{2}-2ac\Gamma ^{2}+\Gamma ^{4}, \\
a &\equiv &k^{2}/(2m_{X})+\mu _{X}-2\Gamma ^{2}/\mu _{C},~ \\
b &\equiv &\mu _{x}-\Gamma ^{2}/\mu _{C}, \\
~c &\equiv &k^{2}/(2m_{C})-\mu _{C}.
\end{eqnarray}%
In the absence of the effective Rabi coupling, i.e., in the case of the
uncoupled waveguiding cores ($\Gamma =0$), branch $\omega _{k}^{-}$ in Eq. (%
\ref{freq}) gives the familiar gapless Bogoliubov-like spectrum,
\begin{equation}
\omega _{k}=\sqrt{{\frac{k^{2}}{2m_{X}}}\left( {\frac{k^{2}}{2m_{X}}}+2\,\mu
_{X}\right) },  \label{freq1}
\end{equation}%
while $\omega _{k}^{+}$ yields
\begin{equation}
\omega _{k}=\left\vert {\frac{k^{2}}{2m_{C}}}-\mu _{C}\right\vert ,
\label{freq2}
\end{equation}%
which may be realized as a gapped spectrum. It is easy to verify that
branches (\ref{freq1}) and (\ref{freq2}) do not intersect, provided that $%
\epsilon _{X}<\mu <\epsilon _{C}$. Notice that for $\Gamma =0$ one has $%
\epsilon _{X}=\mu _{0}^{-}$, see Eq. (\ref{echem}). In the presence of the
effective Rabi coupling ($\Gamma \neq 0$), emulated by the coupling between
the parallel cores, frequencies (\ref{freq}) acquire a finite imaginary part
under the condition of $\mu >\epsilon _{C}$. This means that the uniform
state pertaining to the upper branch of the dispersion relation, i.e., $\mu
^{+}$ in Eq. (\ref{echem}), is always unstable. Instead, for the uniform
state pertaining to the lower branch, characterized by $\mu ^{-}$ in Eq. (%
\ref{echem}), perturbation eigenfrequencies (\ref{freq}) are always real.
Indeed, in the experiments with the EP condensates, only the lower polariton
branch is actually observed.

Dealing with the stability region, in Fig. \ref{fig2} we plot the effective
exciton fraction, $\psi _{0}^{2}/n_{0}$, of the EP-emulating state as a
function of the effective scaled cavity-photon density, $g\phi
_{0}^{2}/(\epsilon _{C}-\epsilon _{X})$, at four values of the scaled
coupling: $\Gamma /(\epsilon _{C}-\epsilon _{X})=0.1,~0.5,1,1.5$ (solid,
dotted, dashed, and dotted-dashed lines, respectively). The figure shows
that the exciton fraction decreases with the increase of $\phi _{0}^{2}$,
while, at a fixed value of $\phi _{0}^{2}$, this fraction slightly grows
with $\Gamma $. In addition, from Fig. \ref{fig2}, and also from Eqs. (\ref%
{cond1}) and (\ref{cond2}), one finds that the uniform state has equal
effective densities of excitons and cavity photons, i.e., $\psi
_{0}^{2}/n_{0}=1/2$, at $g\phi _{0}^{2}=\epsilon _{C}-\epsilon _{X}$.

\subsection{Collective excitations}

The uniform state is stable in the regime where frequencies (\ref{freq}) are
real, i.e., as said above, for the lower branch of the nonlinear dispersion
relation. Here we aim to analyze dispersion relations for collective
excitations on top of the stable uniform state.

In the previous section, it was demonstrated that, in the absence of the
linear coupling ($\Gamma =0$), frequencies (\ref{freq}) split into two
branches, the gapless Bogoliubov-like one (\ref{freq1}), and its gapped
counterpart (\ref{freq2}). In the presence of the coupling ($\Gamma \neq 0$%
), the two branches can be identified: the Bogoliubov-like spectrum, $\omega
_{k}^{-}$, given by Eq. (\ref{freq}), is gapless, i.e. $\omega _{0}^{-}=0$,
while Eq. (\ref{freq}) yields the gapped spectrum, $\omega _{k}^{+}$, with $%
\omega _{0}^{+}\neq 0$.

As mentioned above, the effective exciton and cavity-photon masses are
widely different in the physically relevant setting, $m_{X}\gg m_{C}$,
therefore in many case it is possible to simplify the problem by setting $%
1/m_{X}=0$ \cite{rev-francesca,rev-iacopo}. In this limit case, Eq. (\ref%
{freq}) yields the first-sound velocity $c_{s}$, obtained by the expansion
of the Bogoliubov-like spectrum, $\omega _{k}^{-}$, at small $k$, $\omega
_{k}^{-}\approx c_{s}\,k,$ with
\begin{equation}
c_{s}=\sqrt{{\frac{\epsilon _{C}-\mu }{2m_{C}}}\left[ 1-{\frac{(\mu
_{C}^{2}+\Gamma ^{2})^{2}}{\mu _{C}^{4}+2(\mu _{C}^{2}-\mu _{C}\mu
_{X})\Gamma ^{2}+3\Gamma ^{4}}}\right] }.
\end{equation}%
In the same case, the gap of branch $\omega _{k}^{+}$ is
\begin{equation}
\omega _{0}^{+}=\sqrt{\mu _{C}^{2}+2{\frac{\epsilon _{C}-\epsilon _{X}}{%
\epsilon _{C}-\mu }}\Gamma ^{2}+3{\frac{\Gamma ^{4}}{\mu _{C}^{2}}}}\;.
\end{equation}

\begin{figure}[t]
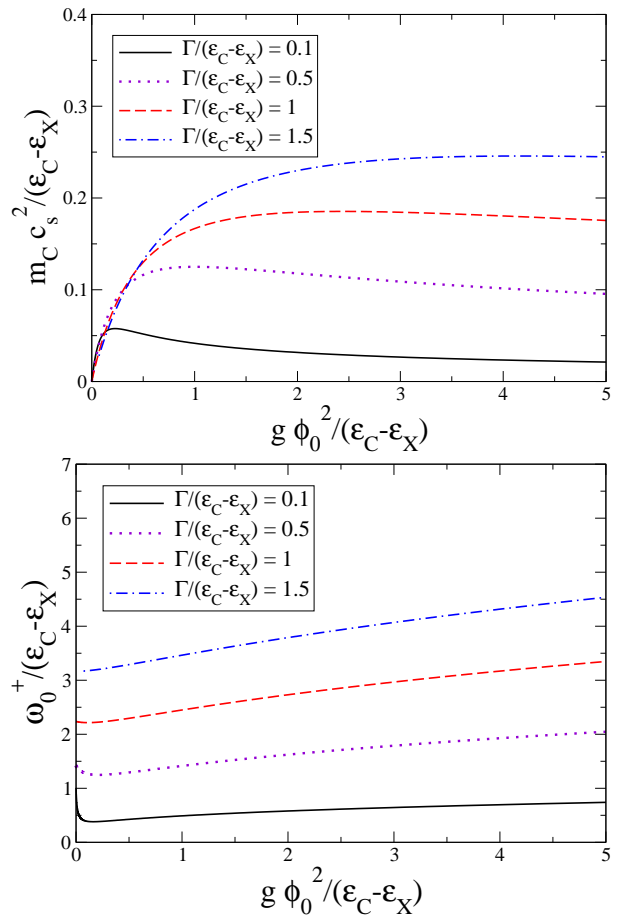

\centerline{\epsfig{file=excitons-f3a.eps,width=8cm,clip=}} %
\centerline{\epsfig{file=excitons-f3b.eps,width=8cm,clip=}}
\caption{(Color online). The top panel: scaled energy $m_{C}c_{s}^{2}/(%
\protect\epsilon _{C}-\protect\epsilon _{X})$ of the first-sound mode, with
the sound speed $c_{s}$, versus the scaled effective cavity-photon density, $%
g\protect\phi _{0}^{2}/(\protect\epsilon _{C}-\protect\epsilon _{X})$. The
bottom panel: scaled energy gap $\protect\omega _{0}^{+}/(\protect\epsilon %
_{C}-\protect\epsilon _{X})$ of the gapped branch versus the scaled
cavity-photon density. In each panel, four curves correspond to different
values of the scaled linear coupling, $\Gamma /(\protect\epsilon _{C}-%
\protect\epsilon _{X})$.}
\label{fig3}
\end{figure}

In Fig. \ref{fig3} we plot the scaled energy, $m_{C}c_{s}^{2}/(\epsilon
_{C}-\epsilon _{X})$, of the first-sound mode (the upper panel), and the
scaled energy gap, $\omega _{0}^{+}/(\epsilon _{C}-\epsilon _{X})$, of the
gapped branch (the lower panel), as functions of the scaled effective
cavity-photon density, $g\phi _{0}^{2}/(\epsilon _{C}-\epsilon _{X})$. Four
curves in each panel correspond to different values of the scaled linear
coupling, $\Gamma /(\epsilon _{C}-\epsilon _{X})$: $0.1,0.5,1,1.5$.

\begin{figure}[t]
\vskip -0.85cm \centerline{%
\epsfig{file=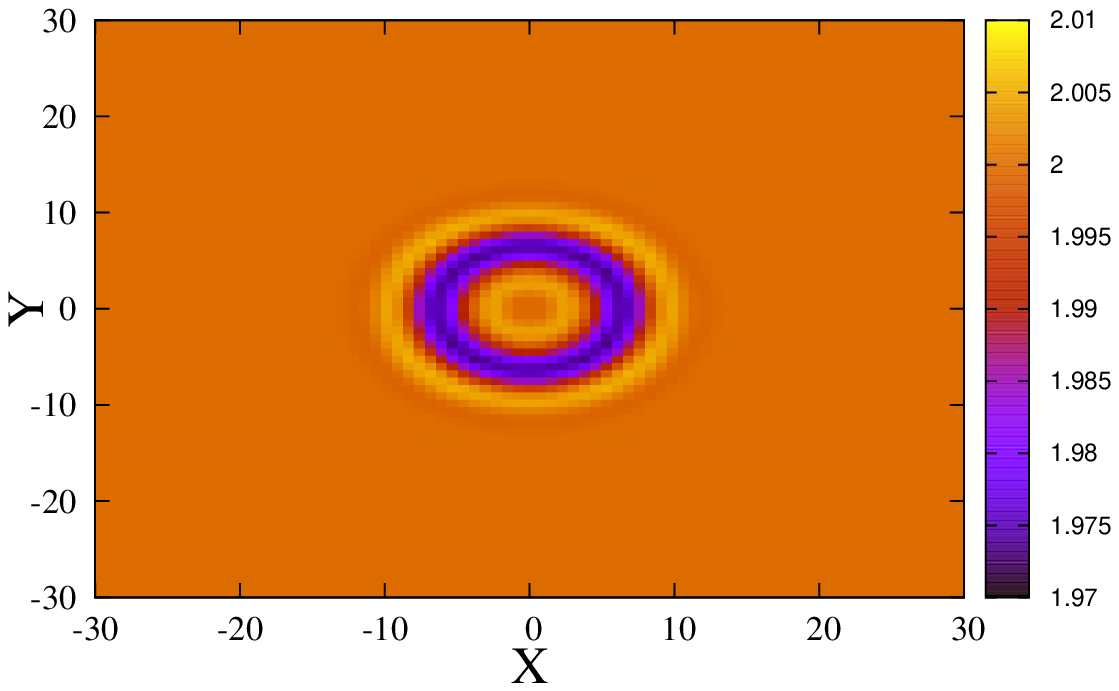,height=6.8cm,width=6.8cm,clip=}} \vskip -1.2cm %
\centerline{\epsfig{file=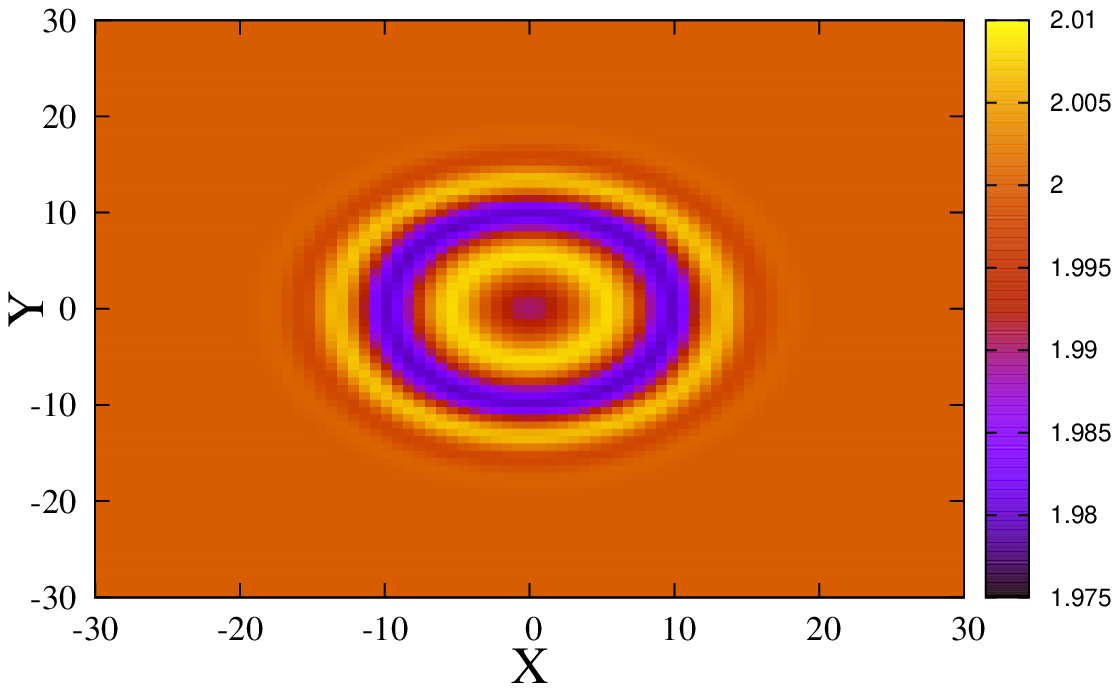,height=6.8cm,width=6.8cm,clip=}} %
\vskip -1.2cm \centerline{%
\epsfig{file=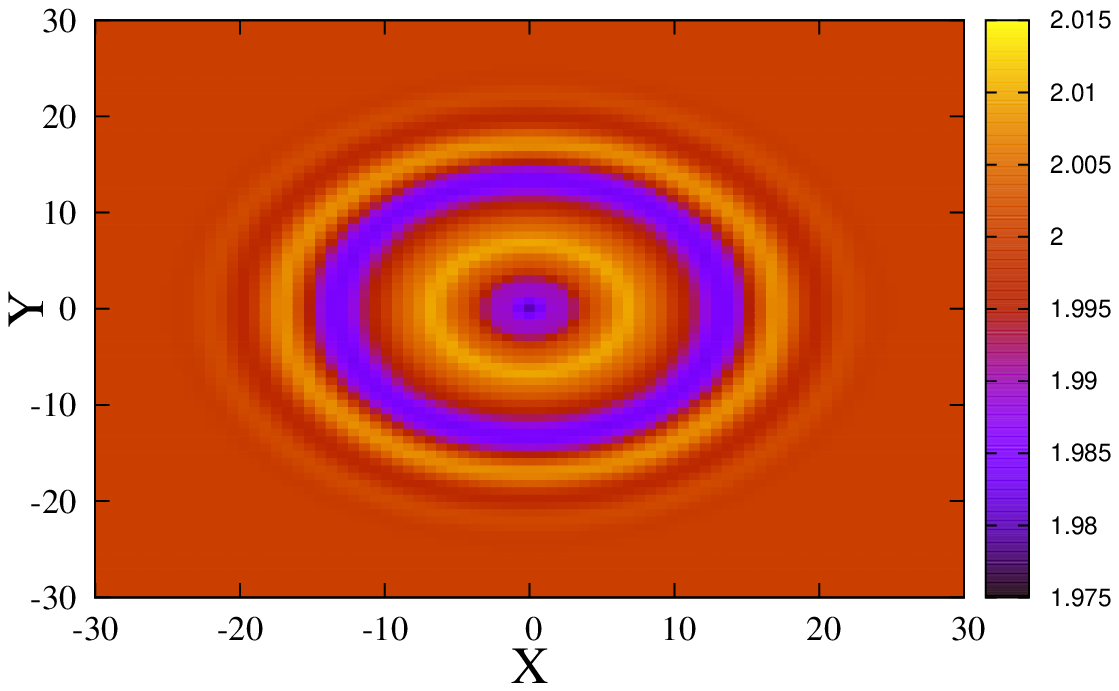,height=6.8cm,width=6.8cm,clip=}} \vskip -0.5cm
\caption{(Color online). The evolution of the initial circular perturbation
in the form of a small hole produced on top of the stable uniform state, per
Eqs. (\protect\ref{initial1}) and (\protect\ref{initial2}). Each panel
displays contour plots of the total density, $n_{0}(x,y,t)$, at fixed values
of the propagation distance (effective time), $t.$ The top, middle, and
bottom panels correspond to $t=9$, $t=15$, and $t=21$, respectively.
Parameters are $\protect\epsilon _{X}=0$, $\protect\epsilon _{C}=1$, $g=1$, $%
\Gamma =0.75$. Initial conditions are taken as in Eqs. (\protect\ref%
{initial1}) and (\protect\ref{initial2}) with $\protect\psi _{0}=1$, $%
\protect\phi _{0}=-1$, $A=0.1$ and $\protect\sigma =2$. }
\label{fig4}
\end{figure}

It is relevant to simulate the evolution of the stable uniform state excited
by a small circular perturbation, which corresponds to an experimentally
relevant situation. In Fig. \ref{fig4} we display the evolution produced by
simulations of Eqs. (\ref{eq1}) and (\ref{eq2}), using a 2D real-time
Crank-Nicolson method with the predictor-corrector element and periodic
boundary conditions \cite{sala-numerics}. For this purpose, we choose the
following initial conditions:
\begin{eqnarray}
\psi (x,t=0) &=&\psi _{0},  \label{initial1} \\
\phi (x,t=0) &=&\phi _{0}+A~e^{-(x^{2}+y^{2})/\sigma ^{2}},  \label{initial2}
\end{eqnarray}%
where $\psi _{0}$ and $\phi _{0}$ are solutions of Eqs. (\ref{cond1}) and (%
\ref{cond2}), while $A$ and $\sigma $ are parameters of the perturbation,
which represents a small circular hole. We here set $\epsilon _{X}=0$, $%
\epsilon _{C}=1$, $g=1$, $\Gamma =0.75$, $m_{X}=1000$ and $m_{C}=1$. Figure %
\ref{fig4} displays the spatial profile of the perturbed condensate density,
$n_{0}(x,t)=|\psi (x,t)|^{2}+|\phi (x,t)|^{2}$, at different values of the
propagation distance (time, in terms of EP), $t=9$, $t=12$, $t=21$, for
initial perturbation (\ref{initial2}) with $A=0.1$ and $\sigma =2$. In the
case shown in Fig. \ref{fig4} the unperturbed amplitudes are $\psi _{0}=1$
and $\phi _{0}=-1$, which correspond in Fig. \ref{fig1} to $\mu =0.25$ (the
lower branch). As observed in Fig. \ref{fig4}, the initial perturbation
produces a circular pattern which expands with a radial velocity close to
the speed of sound, $c_{s}$ (for further technical details, see Ref. \cite%
{sala-shock}).

We have also simulated the evolution of unstable configurations -- for
instance, with unperturbed amplitudes $\psi _{0}=1$ and $\phi _{0}=1$, which
correspond to $\mu =1.75$, i.e., the upper branch in Fig. \ref{fig1}. The
evolution is initially similar to that shown in Fig. \ref{fig4}, but later a
completely different behavior is observed, with the formation of several
circles whose amplitude strongly grows in the course of the unstable
evolution.

\section{The uniform condensate with a phase gradient}

We now analyze the existence and stability of a uniform state with a
phase gradient (effective superflow), which corresponds to setting
\begin{equation}
\left\{ {\psi }(\mathbf{r},t),{\phi }(\mathbf{r},t)\right\} =\left\{ \psi
_{0},\phi _{0}\right\} e^{i(\mathbf{q}\cdot \mathbf{r}-\mu t)}
\end{equation}%
in Eqs. (\ref{eq1}) and (\ref{eq2}), where $\mathbf{q}=(q_{x},q_{y})$ is the
wave vector of the gradient, with $\psi _{0}$ and $\phi _{0}$ given by
\begin{eqnarray}
\psi _{0}^{2} &=&{\frac{1}{g}}\left( {\tilde{\mu}}_{X}-{\frac{\Gamma ^{2}}{{%
\tilde{\mu}}_{C}}}\right) \;,  \label{cond1-grad} \\
\phi _{0} &=&-{\frac{\Gamma }{{\tilde{\mu}}_{C}}}\,\psi _{0}~,
\label{cond2-grad}
\end{eqnarray}%
where
\begin{equation}
{\tilde{\mu}}_{X}=\mu -\epsilon _{X}-{\frac{q^{2}}{2m_{X}}}~,~\ {\tilde{\mu}}%
_{C}=\mu -\epsilon _{C}-{\frac{q^{2}}{2m_{C}}}\;.
\end{equation}%
Following the same procedure as developed in the previous section, we derive
the quartic dispersion equation,
\begin{equation}
\alpha _{4}\omega _{\mathbf{k}}^{4}+\alpha _{3}\omega _{\mathbf{k}%
}^{3}+\alpha _{2}\omega _{\mathbf{k}}^{2}+\alpha _{1}\omega _{\mathbf{k}%
}+\alpha _{0}=0\;,  \label{bestiale}
\end{equation}%
for frequencies $\omega _{k}$ of small excitations on top of the uniform
state. Here we define
\begin{eqnarray}
\alpha _{0} &=&a_{1}a_{2}c_{1}c_{2}-b^{2}c_{1}c_{2}-a_{1}c_{1}\Gamma
^{2}-a_{2}c_{2}\Gamma ^{2}+\Gamma ^{4}\;,  \notag \\
\alpha _{1}
&=&a_{1}a_{2}c_{1}-b^{2}c_{1}-a_{1}a_{2}c_{2}+b^{2}c_{2}+a_{1}c_{1}c_{2}
\notag \\
&-&a_{2}c_{1}c_{2}+a_{1}\Gamma ^{2}-a_{2}\Gamma ^{2}\;,  \notag \\
\alpha _{2} &=&-a_{1}a_{2}+b^{2}+a_{1}c_{1}-a_{2}c_{1}-a_{1}c_{2}  \notag \\
&+&a_{2}c_{2}-c_{1}c_{2}-2\Gamma ^{2}\;,  \notag \\
\alpha _{3} &=&-a_{1}+a_{2}-c_{1}+c_{2}\;,  \notag \\
\alpha _{4} &=&1\;,  \notag
\end{eqnarray}%
\begin{eqnarray}
a_{1} &=&k^{2}/(2m_{X})+{\tilde{\mu}}_{X}-2\Gamma ^{2}/{\tilde{\mu}}_{C}+%
\mathbf{q}\cdot \mathbf{k}/m_{X}\;,  \notag \\
a_{2} &=&k^{2}/(2m_{X})+{\tilde{\mu}}_{X}-2\Gamma ^{2}/{\tilde{\mu}}_{C}-%
\mathbf{q}\cdot \mathbf{k}/m_{X}\;,  \notag \\
b &=&{\tilde{\mu}}_{X}-\Gamma ^{2}/{\tilde{\mu}}_{C}\;,  \notag \\
c_{1} &=&k^{2}/(2m_{C})+{\tilde{\mu}}_{C}+\mathbf{q}\cdot \mathbf{k}/m_{C}\;,
\notag \\
c_{2} &=&k^{2}/(2m_{C})+{\tilde{\mu}}_{C}-\mathbf{q}\cdot \mathbf{k}/m_{C}\;.
\notag
\end{eqnarray}

\begin{figure}[t]
\centerline{\epsfig{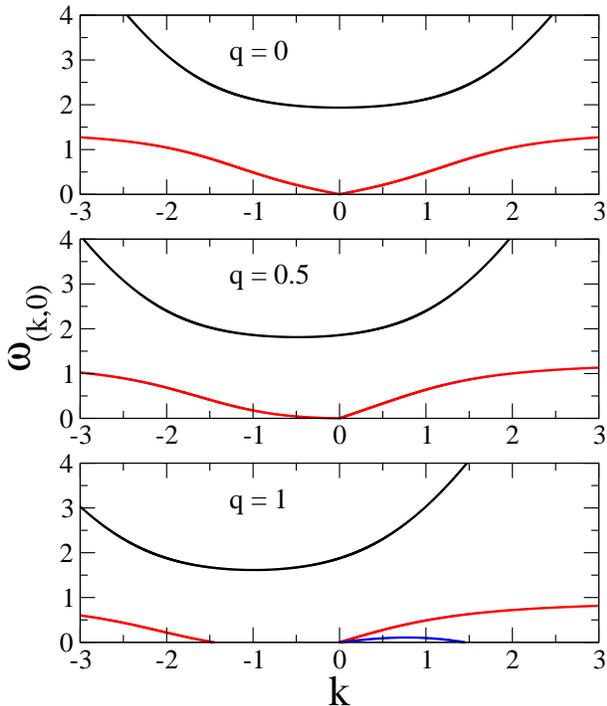}}
\caption{(Color online). Frequencies of small excitations $\protect\omega %
_{(k,0)}$ above the stable uniform state with wave vector $\mathbf{q}=(q,0)$
of the phase flux. Parameters are $\protect\epsilon _{X}=0$, $\protect%
\epsilon _{C}=1$, $g=1$, $\Gamma =0.75$, $m_{X}=1000$, $m_{C}=1$, and $%
\protect\mu =0.25$. Each line corresponds to a different solution (branch)
of Eq. (\protect\ref{bestiale}).}
\label{fig5}
\end{figure}

\begin{figure}[t]
\centerline{\epsfig{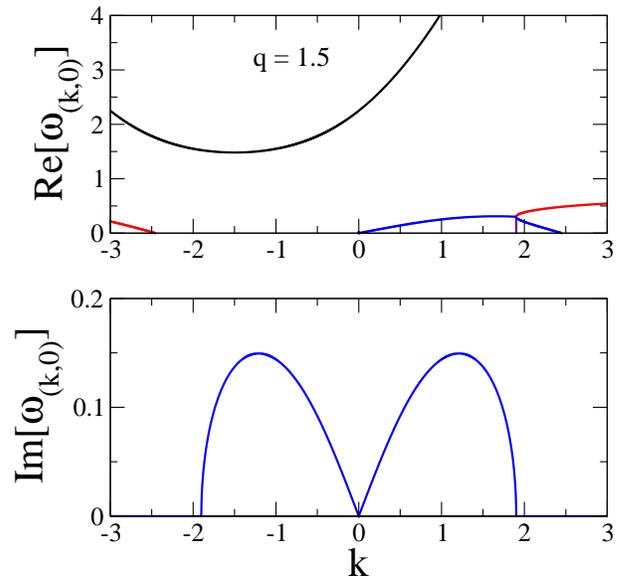}}
\caption{(Color online). Real and imaginary parts, $\mathrm{Re}[\protect%
\omega _{k}]$ and $\mathrm{Im}[\protect\omega _{k}]$, of excitation
frequencies $\protect\omega _{k}$ on top of the unstable uniform state with
wave vector $\mathbf{q}=\left( 1.5,0\right) $ of the phase flux. Parameters
are $\protect\epsilon _{X}=0$, $\protect\epsilon _{C}=1$, $\protect\gamma =1$%
, $\Gamma =0.75$, $m_{X}=1000$, $m_{C}=1$, and $\protect\mu =0.25$. Each
line corresponds to a different solution (branch) of Eq. (\protect\ref%
{bestiale}).}
\label{fig6}
\end{figure}

In the absence of the linear coupling ($\Gamma =0$), Eq. (\ref{bestiale})
gives the $\mathbf{q}$-dependent gapless Bogoliubov-like spectrum,
\begin{equation}
\omega _{\mathbf{k}}={\frac{\mathbf{q}\cdot \mathbf{k}}{m_{X}}}\pm \sqrt{{%
\frac{k^{2}}{2m_{X}}}\left( {\frac{k^{2}}{2m_{X}}}+2\,\left( \mu _{X}-{\frac{%
q^{2}}{2m_{X}}}\right) \right) },
\end{equation}%
and the $\mathbf{q}$-dependent gapped one,
\begin{equation}
\omega _{\mathbf{k}}={\frac{\mathbf{q}\cdot \mathbf{k}}{m_{C}}}\pm
\left\vert {\frac{k^{2}+q^{2}}{2m_{C}}}-\mu _{C}+{\frac{q^{2}}{2m_{C}}}%
\right\vert .
\end{equation}%
In the presence of the linear coupling ($\Gamma \neq 0$) one must solve Eq. (%
\ref{bestiale}) numerically. We direct the $x$ axis along $\mathbf{q}$,
hence $\mathbf{q}=(q,0)$. In Fig. \ref{fig5} we plot frequencies $\omega
_{(k,0)}$ of longitudinal perturbations, with wave vector $\mathbf{k}=\left(
k,0\right) $, for three different values of the flux wavenumber, $q$. For
these values of $q$, the imaginary part of frequencies $\omega _{(k,0)}$ is
zero, hence the state is stable. In the first two panels of Fig. \ref{fig5}
(with $q=0$ and $q=0.5$) one clearly sees gapped and gapless modes, which
are (approximately) symmetric for $k>0$ and $k<0$, which corresponds to
excitation waves moving in opposite directions with equal speeds. By
increasing $q$ one reaches the Landau critical wavenumber, $q_{L}\simeq 0.72$%
, at which the gapless mode has zero frequency at a finite value of $k$, and
above which there is a finite range of $k$'s where two gapless modes
propagate in the same direction, i.e., the phase velocity, $\omega
_{(k,0)}/k $, has the same sign for both the modes. This is shown in the
lower panel of Fig. \ref{fig5} (for $q=1$), where, according to the Landau
criterion, the system is not fully superfluid \cite{book-leggett}. A further
increase of $q$ leads to the dynamical instability of the gapless modes
through the appearance of a nonzero imaginary part of $\omega _{(k,0)}$, as
shown in Fig. \ref{fig6}, where both real and imaginary parts of $\omega
_{(k,0)}$ are displayed for $q=1.5$. Actually, the sound velocities of the
two gapless modes moving in the same direction become equal, so that they
may exchange energy and therefore become unstable, at the critical flux
wavenumber $q_{c}\simeq 1.15$. The results reported in Fig. \ref{fig5} and %
\ref{fig6} are obtained from calculations performed at constant $\mu $. We
have verified that the same phenomenon occurs as well at fixed values of the
total density.

\section{Reduction to the one-dimensional system}

As said above, the 1D lossless EP system may be straightforwardly emulated
by the dual-core optical fiber, in which one core is nonlinear, operating
near the zero-GVD point, while the other one is linear, carrying nonzero
GVD. The accordingly simplified version of Eqs. (\ref{eq1}) and (\ref{eq2})
is written as
\begin{eqnarray}
i{\frac{\partial }{\partial t}}{\tilde{\psi}} &=&\left[ \epsilon _{X}+g\,|{%
\tilde{\psi}}|^{2}\right] {\tilde{\psi}}+\Gamma \ {\tilde{\phi}}\;,
\label{eq1pip} \\
i{\frac{\partial }{\partial t}}{\tilde{\phi}} &=&\left[ -{\frac{1}{2m_{C}}}{%
\frac{\partial ^{2}}{\partial x^{2}}}+\epsilon _{C}+V(x)\right] {\tilde{\phi}%
}+\Gamma \ {\psi }\;,  \label{eq2pip}
\end{eqnarray}%
where the tildes stress the reduction to 1D.

\subsection{Stability of the uniform 1D state}

\begin{figure}[t]
\centerline{\epsfig{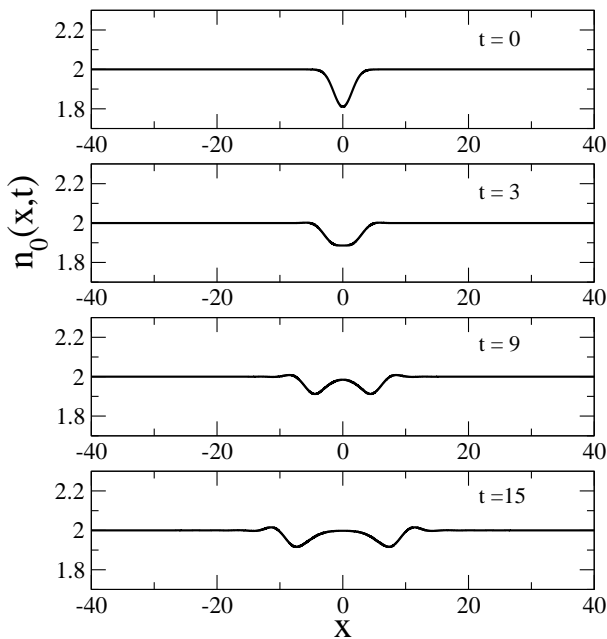}}
\caption{(Color online) The evolution of a small hole produced, as a
perturbation, on top of a stable uniform 1D state. In each panel, density
profiles $n\left( x,t\right) $ are displayed. Parameters are: $\protect%
\epsilon _{X}=0$, $\protect\epsilon _{C}=1$, $g=1$, $\Gamma =0.75$. Initial
conditions are given by Eqs. (\protect\ref{initial1}) and (\protect\ref%
{initial2}) with $\protect\psi _{0}=1$, $\protect\phi _{0}=-1$, $A=0.1$ and $%
\protect\sigma =2$. }
\label{fig7}
\end{figure}

\begin{figure}[t]
\centerline{\epsfig{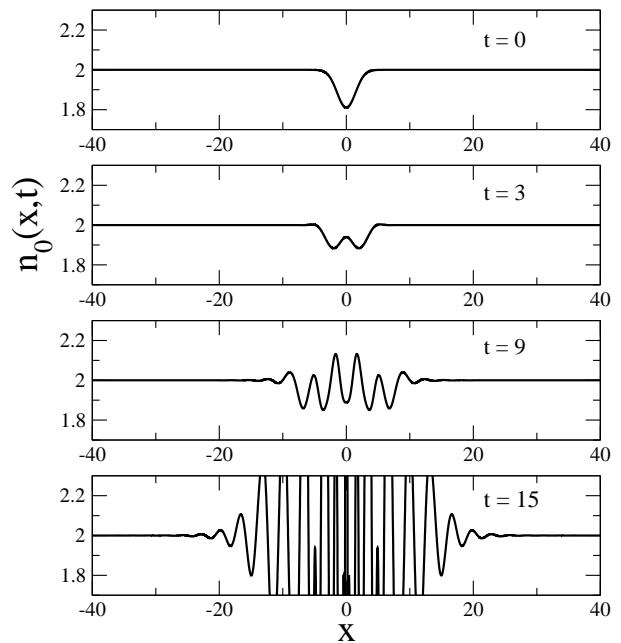}}
\caption{(Color online) The evolution of a small hole produced, as a
perturbation, on top of an unstable 1D state. Parameters are $\protect%
\epsilon _{X}=0$, $\protect\epsilon _{C}=1$, $g=1$, $\Gamma =0.75$. Initial
conditions are given by Eqs. (\protect\ref{initial1}) and (\protect\ref%
{initial2}) with $\protect\psi _{0}=1$, $\protect\phi _{0}=1$, $A=-0.1$ and $%
\protect\sigma =2$.}
\label{fig8}
\end{figure}

The results shown in Figs. \ref{fig1} and \ref{fig2} are also valid in 1D,
taking into regard that $\psi _{0}^{2}$ and $\phi _{0}^{2}$ are now 1D
densities. In Figs. \ref{fig7} and \ref{fig8}, we show the evolution of
small perturbations on top of the uniform stable and unstable states,
respectively. For this purpose, Eqs. (\ref{eq1}) and (\ref{eq2}) were
simulated by means of the 1D real-time Crank-Nicolson algorithm with the use
of the predictor-corrector element \cite{sala-numerics}. The initial
conditions were taken as%
\begin{eqnarray}
{\tilde{\psi}}(x,t=0) &=&\psi _{0}\;,  \label{initial1pip} \\
{\tilde{\phi}}(x,t=0) &=&\phi _{0}+A~e^{-x^{2}/\sigma ^{2}}.
\label{initial2pip}
\end{eqnarray}%
In both Figs. \ref{fig7} and \ref{fig8}, the same parameters of the system
are used: $\epsilon _{X}=0$, $\epsilon _{C}=1$, $g=1$, $\Gamma =0.75$. Both
figures \ref{fig7} and \ref{fig8} display spatial profiles of the condensate
density, $n_{0}(x,t)=|{\tilde{\psi}}(x,t)|^{2}+|{\tilde{\phi}}(x,t)|^{2}$,
at different values of the propagating constant (alias time, in terms of the
EP system), $t=0$, $t=3$, $t=9$, $t=12$, generated by the initial
perturbation (\ref{initial2pip}), with $A=-0.1$ and $\sigma =2$. In Fig. \ref%
{fig7}, the unperturbed amplitudes are $\psi _{0}=1$ and $\phi _{0}=-1$,
which correspond to $\mu =0.25$, i.e., the lower branch in terms of Fig. \ref%
{fig1}, while in Fig. \ref{fig8} the initial amplitudes are $\psi _{0}=1$
and $\phi _{0}=1$, corresponding to $\mu =1.75$ on the upper branch.

As seen in Fig. \ref{fig7}, the initial perturbation hole splits into two
ones traveling in opposite directions with a velocity close to the speed of
sound $c_{s}$ (see further technical details in Ref. \cite{sala-shock}). In
Fig. \ref{fig8}, the dynamics is initially (at $t\leq 6$) similar to that in
Fig. \ref{fig7}, but at $t>6$ it displays a completely different behavior,
namely, formation of strong oscillations, which increase their amplitude in
the course of the evolution, indicating a dynamical instability.

\subsection{One-dimensional dark solitons , and a possibility of the
existence of vortices in the 2D system}

In the case when the 1D uniform background is stable, it is natural to look
for solutions in the form of dark solitons (DSs). As well as in other
systems, the node at the center of the DS is supported by a phase shift of $%
\pi $ between the wave fields at $x\rightarrow \pm \infty $ \cite{Dark} (as
shown below, in the present system the node and the phase shift by $\pi $
exist simultaneously in both fields, $\psi $ and $\phi $).

To demonstrate the possibility of the existence of the DS, we substitute the
general 1D ansatz for stationary solutions into Eqs. (\ref{eq1pip}) and (\ref%
{eq2pip}):
\begin{equation}
\left\{ {\tilde{\psi}}(x,t),{\tilde{\phi}}(x,t)\right\} =\left\{ \Psi
(x),\Phi (x)\right\} \ e^{-i\mu t}\;,  \label{tilde}
\end{equation}%
with functions $\Psi (x)$ and $\Phi (x)$, which may be assumed real, obeying
the coupled stationary equations:
\begin{gather}
\Gamma \,\Phi =\left( \mu -\epsilon _{X}\right) {\Psi }-g\,\Psi ^{3},
\label{Psi} \\
\left( \mu -\epsilon _{C}\right) {\Phi }+\frac{1}{2m_{C}}\Phi ^{\prime
\prime }-\Gamma \ {\Psi }=0,  \label{Phi}
\end{gather}%
where approximation $1/m_{X}=0$ is adopted, and $\Phi ^{\prime \prime
}\equiv d^{2}\Phi /dx^{2}$. It is straightforward to check that Eqs. (\ref%
{Psi}) and (\ref{Phi}) admit a solution with $\Psi (x)=-\Phi (x)$ only if ${%
\Phi }^{\prime \prime }\equiv 0$ for any $x$, i.e., DS solutions do not obey
this constraint.

For the analytical consideration, we assume that the second derivative in
Eq. (\ref{Phi}) may be treated as a small term. Then, an approximate
solution of Eq. (\ref{Phi}) is%
\begin{equation}
\Phi \approx -\frac{\Gamma }{\epsilon _{C}-\mu }\Psi -\frac{\Gamma }{%
2m_{C}\left( \epsilon _{C}-\mu \right) ^{2}}\Psi ^{\prime \prime }\;,
\label{PhiPsi}
\end{equation}%
and the substitution of expression (\ref{PhiPsi}) into Eq. (\ref{Psi}) leads
to the following equation for $\Psi (x)$:
\begin{equation}
-\frac{\Gamma ^{2}}{2m_{C}\left( \epsilon _{C}-\mu \right) ^{2}}\Psi
^{\prime \prime }+g\Psi ^{3}-\left[ \left( \mu -\epsilon _{X}\right) +\frac{%
\Gamma ^{2}}{\left( \epsilon _{C}-\mu \right) }\right] \Psi =0.
\label{single}
\end{equation}%
Equation (\ref{single}) yields a commonly known exact dark-soliton solution:
\begin{align}
\Psi (x)& =\pm \sqrt{\frac{1}{g}\left[ \left( \mu -\epsilon _{X}\right) +%
\frac{\Gamma ^{2}}{\left( \epsilon _{C}-\mu \right) }\right] }  \notag \\
\times & \tanh \left( \frac{1}{\Gamma }\sqrt{m_{C}\left( \epsilon _{C}-\mu
\right) \left[ \left( \mu -\epsilon _{X}\right) \left( \epsilon _{C}-\mu
\right) +\Gamma ^{2}\right] }x\right) \;.  \label{dark}
\end{align}

\begin{figure}[t]
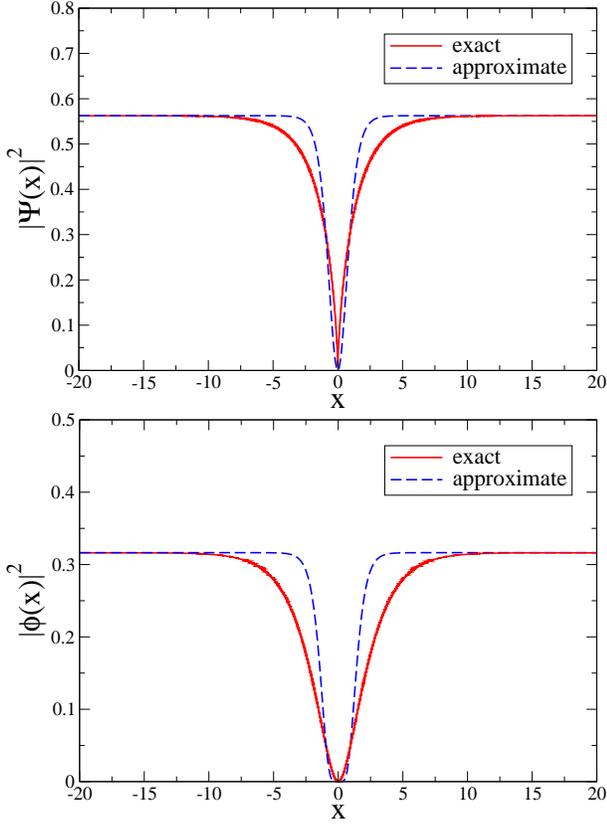

\centerline{\epsfig{file=excitons-f9a.eps,width=8cm,clip=}} %
\centerline{\epsfig{file=excitons-f9b.eps,width=8cm,clip=}}
\caption{(Color online) The dark soliton for $\protect\epsilon _{X}=\protect%
\mu =0$, $\protect\epsilon _{C}=1$, $\protect\gamma =1$, $\Gamma =0.75$. The
exact solution (solid line) is produced by Eq. (\protect\ref{darkmio}), and
its approximate counterpart (dashed line) is given by Eq. (\protect\ref{dark}%
). The top and bottom panels display the excitonic and photonic density
profiles, $|\Psi (x)|^{2}$ and $|\Phi (x)|^{2}$, respectively.}
\label{fig9}
\end{figure}

On the other hand, in the special case of $\mu =\epsilon _{X}$, Eq. (\ref%
{Psi}) can be used to eliminate $\Phi $ in favor of $\Psi $:%
\begin{equation}
\Phi =-\left( g/\Gamma \right) \Psi ^{3},
\end{equation}%
the remaining equation for $\chi \equiv \Psi ^{3}$ being%
\begin{equation}
\frac{1}{2m_{C}}\frac{d^{2}\chi }{dx^{2}}=\left( \epsilon _{C}-\epsilon
_{X}\right) \chi -\frac{\Gamma ^{2}}{g}\chi ^{1/3}.  \label{chi}
\end{equation}%
If $x$ is formally considered as time, Eq. (\ref{chi}) is the Newton's
equation of motion for a particle in an effective external potential,%
\begin{equation}
U_{\mathrm{eff}}(\chi )=\frac{3\Gamma ^{2}}{4g}\chi ^{4/3}-\frac{1}{2}\left(
\epsilon _{C}-\epsilon _{X}\right) \chi ^{2}.  \label{U}
\end{equation}%
It is obvious that this potential gives rise to a heteroclinic trajectory
which connects two local maxima of potential (\ref{U}), $\chi _{0}=\pm \left[
\Gamma ^{2}/\left( g\left( \epsilon _{C}-\epsilon _{X}\right) \right) \right]
^{3/2}$. An implicit analytical form of $\chi (x)$ for the corresponding
solution given by
\begin{equation}
x=\int_{0}^{\chi (x)}d\xi \sqrt{\frac{2}{\left(
4s_{1}^{3}/27s_{0}^{2}\right) +s_{0}\xi ^{2}-s_{1}\xi ^{4/3}}}\;,
\label{darkmio}
\end{equation}%
where $s_{0}=m_{C}(\epsilon _{C}-\epsilon _{X})$ and $s_{1}=3m_{C}\Gamma
^{2}/(2g)$. From $\chi (x)$ one obtains $\Psi (x)=\left[ \chi (x)\right]
^{1/3}$ and $\Phi (x)=-(g/\Gamma )\chi (x)$. In Fig. \ref{fig9} we compare
the exact implicit DS solution (solid line), produced by Eq. (\ref{darkmio}%
), and its approximate counterpart (dashed line) given by Eq. (\ref{dark}).

Finally, getting back to the full 2D system of Eqs. (\ref{eq1}) and (\ref%
{eq2}), and substituting there $\left\{ {\psi }(\mathbf{r},t),{\phi }(%
\mathbf{r},t)\right\} =\left\{ \Psi (\mathbf{r}),\Phi (\mathbf{r})\right\}
e^{-i\mu t}$, cf. Eq. (\ref{tilde}), we note that the existence of 2D
vortices can be predicted by means of the approximation similar to that in
Eq. (\ref{PhiPsi}), i.e.,%
\begin{equation}
\Phi \approx -\frac{\Gamma }{\epsilon _{C}-\mu }\Psi -\frac{\Gamma }{%
2m_{C}\left( \epsilon _{C}-\mu \right) ^{2}}\nabla _{\perp }^{2}\Psi .
\end{equation}%
The substitution of this into the stationary version of Eq. (\ref{eq1}) with
$1/m_{X}=0$ yields
\begin{gather}
-\frac{\Gamma ^{2}}{2m_{C}\left( \epsilon _{C}-\mu \right) ^{2}}\nabla
_{\perp }^{2}\Psi +g\left\vert \Psi \right\vert ^{2}\Psi  \notag \\
-\left[ \left( \mu -\epsilon _{X}\right) +\frac{\Gamma ^{2}}{\left( \epsilon
_{C}-\mu \right) }\right] \Psi =0,  \label{2Dsingle}
\end{gather}%
cf. Eq. (\ref{single}). It is the usual 2D nonlinear Schr\"{o}dinger
equation with the self-defocusing nonlinearity, which gives rise to commonly
known vortex states \cite{vortices}.

\section{Conclusions}

The objective of this work is to propose the dual-core optical waveguide,
with one linear and one dispersive cores, as an emulator for the EP
(exciton-polariton) system in the lossless limit, which is not currently
achievable in semiconductor microcavities. In terms of this model, the first
fundamental issue is the MI\ (modulational instability) of the uniform
state. As might be expected, it is found that the uniform states
corresponding to the upper and lower branches of the nonlinear dispersion
relation are, respectively, unstable and stable. This analytical result is
confirmed by direct simulations, which demonstrate the evolution of
localized perturbations on top of stable and unstable backgrounds. The
excitation modes supported by the stable background are analyzed too,
demonstrating two gapless and gapped branches in the spectrum. The stability
investigation was generalized for the uniform background with the phase
flux, demonstrating that the lower-branch state loses the stability at the
critical value of the flux wavenumber. Finally, approximate and exact
analytical solutions for stable dark solitons supported by the 1D setting
are produced too.

The analysis may be extended in other directions. In particular, a
challenging problem is an accurate investigation of 2D vortices, the
existence of which is suggested by the approximate equation (\ref{2Dsingle}).

\section*{Acknowledgments}

The authors acknowledge for partial support Universit\`{a} di Padova (grant
No. CPDA118083), Cariparo Foundation (Eccellenza grant 11/12), and MIUR
(PRIN grant No. 2010LLKJBX). The visit of B.A.M. to Universit\`{a} di Padova
was supported by the Erasmus Mundus EDEN grant No. 2012-2626/001-001-EMA2.
L.S. thanks F.M. Marchetti for useful e-discussions.

\end{document}